# Variations of saturation vapor pressure and evaporation rate of liquids with their vaporization enthalpy

*Xuefeng Xu,*[\*a] *Chengzhi Yu,*[a] *and Liran Ma*[†b]

[a] School of Technology, Beijing Forestry University, Beijing 100083, China
[b] State Key Laboratory of Tribology in Advanced Equipment, Tsinghua University, Beijing 100084, China

**ABSTRACT:** The phase transition enthalpy of condensed materials can be altered by factors such as electric fields, and such variations in turn affect physical and chemical behaviors including phase equilibrium. However, due to the neglect of non-expansion work, the Clapeyron equation does not account for the effect of changes in phase transition enthalpy on equilibrium. In this paper, by analyzing the electric-field-induced changes in phase transition enthalpy and incorporating the non-expansion work performed on the system, we extended both the Clapeyron and Clausius–Clapeyron equations to explicitly include variations in phase transition enthalpy. Building upon these extensions, analytical expressions for the vapor pressure of liquids and for the total evaporation rate of sessile liquid droplets as functions of the change in vaporization enthalpy have been derived, showing that an approximately 1% decrease in vaporization enthalpy of a liquid can cause an increase of about 20% in its vapor pressure and an almost twofold increase in its evaporation rate. The theoretical predictions by the present equations were compared and found to be in good agreement with the experimental observations. The present work may contribute to the body of knowledge concerning phase equilibrium or phase transition and may have applications in a wide range of scientific and industrial processes.

---

[\*] Corresponding author. E-mail: xuxuefeng@bjfu.edu.cn

[†] Corresponding author. E-mail: maliran@tsinghua.edu.cn

# 1. Introduction

The cohesive energy $E_{coh}$ of condensed-matter systems is defined as the average energy required to separate all constituent molecules to an infinite distance from each other without altering the internal energy of the individual molecules [1]. $E_{coh}$ corresponds to the change in internal energy during the transition from condensed matter to the vapor phase, i.e., the internal energy of vaporization $\Delta_{vap}U$ for liquids or the internal energy of sublimation $\Delta_{sub}U$ for solids [2]. Assuming that the vapor phase behaves as an ideal gas, the molar cohesive energy $E_{coh,m}=\Delta_{vap}H_m - RT$ for liquid state and $E_{coh,m}=\Delta_{sub}H_m - RT$ for solid state, where $\Delta_{vap}H_m$ is the molar enthalpy of vaporization, $\Delta_{sub}H_m$ is the molar enthalpy of sublimation, $R$ denotes the universal gas constant, and $T$ represents the absolute temperature [1-5]. These imply that, during an isothermal process, the change in these phase transition enthalpies is exactly equal to the change in the cohesive energy.

The phase transition enthalpies or the cohesive energy of condensed materials can be modified by various factors, including the number of $CH_2$ groups in molecule of liquid hydrocarbons [1,3], the chain length of molecule in Lennard-Jones liquids [6], the molar refraction [7] and the molar volume [8] of ionic liquids, and the addition of alkoxy chains to ionic liquids [9]. Moreover, experiments and simulations have found that applying an electric field can also alter the phase transition enthalpies or the cohesive energy. Molecular dynamics simulations show that increasing the electric field strength leads to a corresponding rise in the cohesive energy of liquid Nitromethane by inducing an ordering effect on the structure of liquid [10]. The extensively studied electrocaloric effect refers to the entropy change observed in ferroelectric materials upon application of an electric field. This phenomenon further implies that the phase transition enthalpies of both solid-state materials [11-14] and liquid crystals [15,16] can be altered by an electric field which can induce a change in the dipolar state in a dielectric material from a disordered to an ordered one.

As fundamental thermodynamic properties, variations in the phase transition enthalpies or in the cohesive energy affect a wide range of physical and chemical behaviors, such as solubility [1, 3], fluidity [8], the equation of state [17]. Furthermore, the Clapeyron equation demonstrates that the enthalpy of phase transition also exerts a significant influence on phase equilibrium [18]. However, although the equation includes the parameter of phase transition enthalpy, it merely establishes a quantitative relationship between pressure and temperature variations under two-phase equilibrium, offering no insight into how changes in phase transition enthalpy influences that equilibrium. This limitation stems from the neglect of non-expansion work during the equation's derivation.

In this work, the change in enthalpy of a system induced by electric fields was analyzed, and the Clapeyron equation and the Clausius–Clapeyron equation were extended to incorporate changes in phase transition enthalpy by considering non-expansion work performed on the system. From these extended equations, analytical expressions were derived for the saturation vapor pressure of liquids and the evaporation rate of sessile liquid droplets as functions of variations in the phase transition enthalpy. Experimental observations of water droplets under electric fields were then performed and showed strong agreement with the theoretical predictions from the proposed equations.

## 2. Results and Discussion

### 2.1 An extended Clapeyron equation

Here, the change in enthalpy of system under the application of an electric field was first analyzed. When an electric field is applied to dielectric materials, it will increase the dipolar ordering in the materials [11-16]. The transition of the dipolar state from a disordered to an ordered one results from the non-expansion work performed by the electric field on the material and will induce a change of enthalpy of the material. For an infinitesimal change in the state of the system during which non-expansion work $dw_{\text{non-exp}}$ has been done on the system by the electric field, the change in enthalpy of the system $dH = TdS + Vdp + dw_{\text{non-exp}}$, where $S$ is the entropy of the system, $V$ the volume, and $p$ the pressure [18]. For bulk systems, the temperature change of the systems when adiabatically applying electrical fields is typically very small [11,12,15]. This means that, under isothermal conditions, the energy transferred as heat to the system, i.e., the term *TdS*, can reasonably be neglected, thereby simplifying the equation to $dH = dw_{\text{non-exp}}$ at constant pressure.

During the derivation of the Clapeyron equation, the fundamental thermodynamic equation $dG = Vdp - SdT$, which gives the variation of Gibbs free energy *G* with pressure *p* and temperature *T*, was applied. Because this equation is strictly valid only for a closed system doing no non-expansion work, the Clapeyron equation itself is applicable exclusively to such a system [18]. When non-expansion work is also performed on the system by electric field, the variation of *G* can be expressed as $dG = Vdp - SdT + dw_{\text{non-exp}} = Vdp - SdT + dH$.

Supposing that two phases *α* and *β* of a substance are in equilibrium in the system, their molar Gibbs energy $G_m$ are equal. For an infinitesimal change in the state of the system in such a way that the phases remain in equilibrium, the change in $G_m$ of phase *α* must be the same as that of phase *β*, i.e., $V_m(\alpha)dp - S_m(\alpha)dT + dH_m(\alpha) = V_m(\beta)dp - S_m(\beta)dT + dH_m(\beta)$, where $S_m(\alpha)$ and $S_m(\beta)$ are the molar entropies of the two phases, $V_m(\alpha)$ and $V_m(\beta)$ are their molar volumes, and $H_m(\alpha)$ and $H_m(\beta)$ are their molar enthalpies. This leads to the change in the pressure:

$$dp = \frac{\Delta_{\text{trs}} S_m}{\Delta_{\text{trs}} V_m} dT - \frac{1}{\Delta_{\text{trs}} V_m} d(\Delta_{\text{trs}} H_m) \tag{1}$$

where $\Delta_{\text{trs}} V_m = V_m(\beta) - V_m(\alpha)$, $\Delta_{\text{trs}} S_m = S_m(\beta) - S_m(\alpha)$, $\Delta_{\text{trs}} H_m = H_m(\beta) - H_m(\alpha)$ are the molar volume, the molar entropy, and the molar enthalpy of phase transition respectively. Equation (1) is an extended Clapeyron equation, which takes the changes in phase transition enthalpy into account.

**2.2 An extended Clausius–Clapeyron equation for liquid-vapor equilibrium**

For liquid-vapor equilibrium, the molar entropy of vaporization $\Delta_{\text{vap}} S_m$ at the temperature *T* is equal to $\frac{\Delta_{\text{vap}} H_m}{T}$, and thus, the extended Clapeyron equation can be written as $dp = \frac{\Delta_{vap}H_m}{T\Delta_{vap}V_m}dT - \frac{1}{\Delta_{vap}V_m}d(\Delta_{vap}H_m)$. Assuming that the vapor is an ideal gas, its molar volume is given by $V_{v,m} = RT/p$. Considering also that the molar volume of the liquid is typically much smaller than that of the vapor (i.e., $V_{l,m} \ll V_{v,m}$), $\Delta_{vap}V_m \approx V_{v,m} = RT/p$. Incorporating this approximation into the above expression, we obtain the extended Clausius–Clapeyron equation as follows:

$$dp = \frac{\Delta_{vap}H_m p}{RT^2}dT - \frac{p}{RT}d(\Delta_{vap}H_m) \qquad (2)$$

or

$$d\ln p = \frac{\Delta_{vap}H_m}{RT^2}dT - \frac{1}{RT}d(\Delta_{vap}H_m) \qquad (3)$$

**2.3 Variation of vapor pressure with vaporization enthalpy**

We consider a liquid-vapor equilibrium system. Initially, the system temperature is $T_0$, the molar enthalpy of vaporization is $\Delta_{vap}H_m^0$, and the corresponding saturation vapor pressure is $p_0$. When the system temperature changes to $T$ and the molar enthalpy of vaporization simultaneously changes to $\Delta_{vap}H_m$, the equilibrium vapor pressure becomes $p$. Here, we design a reversible two-stage pathway to realize these changes: first, keep the molar vaporization enthalpy of liquid constant at $\Delta_{vap}H_m^0$ while the system temperature changes reversibly from $T_0$ to $T$; then, keep the system temperature fixed at $T$ while the molar vaporization enthalpy of liquid changes reversibly from $\Delta_{vap}H_m^0$ to $\Delta_{vap}H_m$ by applying an electric field. Integrating the extended Clausius–Clapeyron equation along this two-step path yields:

$$p = p_0 e^{\frac{\Delta_{vap}H_m^0}{RT_0} - \frac{\Delta_{vap}H_m}{RT}} \qquad (4)$$

Under isothermal conditions, the variation in saturation vapor pressure with respect to the change in molar vaporization enthalpy can be described as:

$$p = p_0 e^{-\frac{\Delta_{vap}H_m - \Delta_{vap}H_m^0}{RT}} \qquad (5)$$

The equation (5) clearly demonstrates that a decrease in the vaporization enthalpy of a liquid leads to an increase in the saturation vapor pressure of the liquid. This is because a reduction in the vaporization enthalpy or the cohesive energy reflects weakened intermolecular interactions within the liquid, thereby facilitating the escape of molecules from the liquid surface into the vapor phase. Conversely, an increase in the vaporization enthalpy or the cohesive energy strengthens these interactions, hindering molecular escape and thus lowering the saturation vapor pressure. Calculations reveal that a change of just 0.5 kJ/mol in the vaporization enthalpy can induce a variation of up to approximately 20% in the saturation vapor pressure of the liquid (see Figure 1a). Given that the vaporization enthalpy of water at 25°C is 43.99 kJ/mol [19], the results indicate that relatively small variations in vaporization enthalpy can lead to substantial changes in saturation vapor pressure.

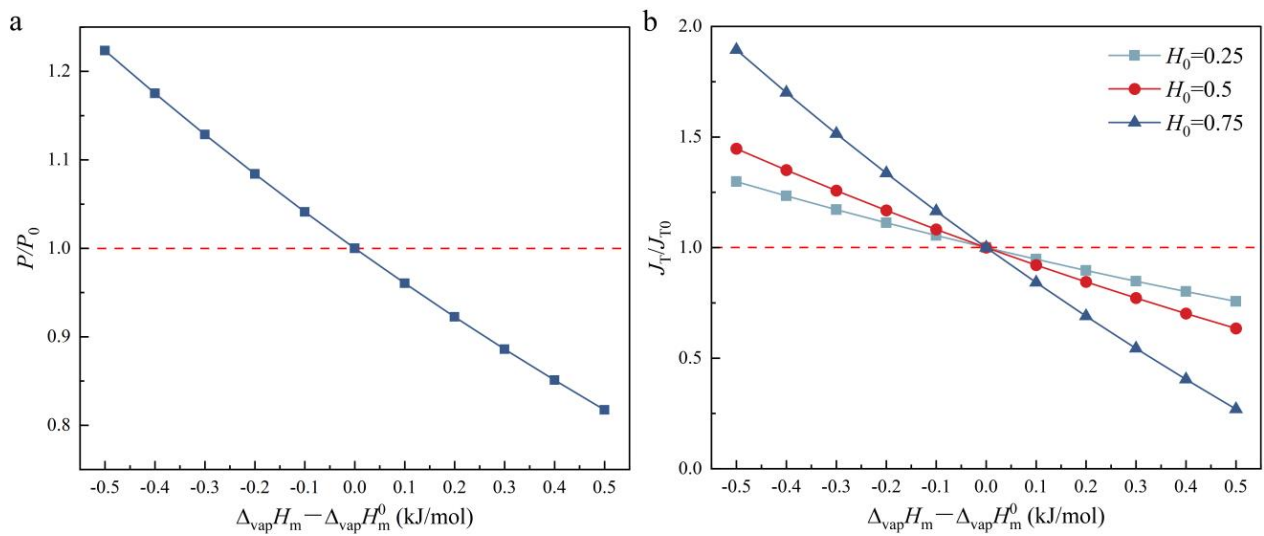

**Figure 1. (a)** Variation in the saturation vapor pressure of a liquid and **(b)** variations in the evaporation rate of liquid droplets as functions of changes in molar vaporization enthalpy, calculated by using equations (5) and (7), respectively.

## 2.4 Variation of droplet evaporation rate with vaporization enthalpy

Here, we consider a slowly evaporating liquid droplet with a contact line radius of $R$ and a contact angle of $\theta$ resting on a flat substrate. Due to the small Bond number and capillary number, the droplet shape can be considered as a spherical cap [20,21]. Along the surface of the droplet, the temperature can be considered to be constant at $T$ and the vapor concentration is also constant at $c_S$ which is the saturated vapor concentration of the liquid at temperature $T$ [20,21]. The slow evaporation process can be regarded as diffusion-controlled and quasi-steady, implying that the vapor concentration $c$ in the surrounding atmosphere satisfies Laplace's equation [20,21]. By choosing scaling factors for the nondimensionalization as follows: $\tilde{c} = \dfrac{c - c_S}{c_S(1-H)}$, where $H$ is the relative humidity of the ambient air, the governing equations and the boundary conditions for the evaporation process are:

$$\tilde{\nabla}^2 \tilde{c} = 0 \quad \text{for} \quad \tilde{z} \geq \tilde{h}_L(\tilde{r}) \tag{6.1}$$

$$\tilde{c} = -1 \quad \text{for} \quad \tilde{z} = \infty, \tilde{r} = \infty \tag{6.2}$$

$$\tilde{c} = 0 \quad \text{for} \quad \tilde{z} = \tilde{h}_L(\tilde{r}), \quad \tilde{r} \leq 1 \tag{6.3}$$

$$\frac{\partial \tilde{c}}{\partial \boldsymbol{n}} = 0 \quad \text{for} \quad \tilde{z} = 0, \quad \tilde{r} > 1 \tag{6.4}$$

Where $\tilde{\nabla}^2 = \dfrac{\partial^2}{\partial^2 \tilde{r}} + \dfrac{1}{\tilde{r}}\dfrac{\partial}{\partial \tilde{r}} + \dfrac{\partial^2}{\partial^2 \tilde{z}}$, $\tilde{r} = \dfrac{r}{R}$, $\tilde{z} = \dfrac{z}{R}$, $\tilde{h}_L(\tilde{r}) = h_L(r)/R = \sqrt{(1/\sin\theta)^2 - \tilde{r}^2} - 1/\tan(\theta)$ is the height of the droplet, and $\boldsymbol{n}$ is the unit normal. It can be easily seen from the equations that the nondimensional evaporation flux, defined as $\tilde{J}(\tilde{r}) = -\tilde{\nabla}\tilde{c} \cdot \boldsymbol{n} = \dfrac{J(r)R}{Dc_S(1-H)}$, where $J(r) = -D\nabla c \cdot \boldsymbol{n}$ is the evaporation flux from the droplet surface, and the nondimensional total evaporation rate, defined as

$$\tilde{J}_T = \int_0^1 2\pi \tilde{r} \tilde{J}(\tilde{r}) \sqrt{1 + (\frac{\partial \tilde{h}}{\partial \tilde{r}})^2}\, d\tilde{r} = \frac{J_T}{RDc_S(1-H)}, \quad \text{where} \quad J_T = \int_0^R 2\pi r J(r)\sqrt{1 + (\frac{\partial h(r,t)}{\partial r})^2}\, dr \quad \text{is the total}$$

evaporation rate from the whole droplet surface, are identical for droplets with the same contact angle.

We consider the evaporation of two droplets with the same contact angle $\theta$ and base radius $R$ under the identical temperature. The two droplets are composed of the same liquid but are subjected to different electric fields. One droplet has a molar vaporization enthalpy $\Delta_{vap}H_m^0$ and the corresponding saturated vapor concentration, relative humidity and total evaporation rate are $c_{S0}$, $H_0$, and $J_{T0}$, respectively, while another droplet has a molar vaporization enthalpy $\Delta_{vap}H_m$, with corresponding saturated vapor concentration $c$, relative humidity $H$, and evaporation rate $J_T$. From equation (5), we obtain the relation between the vapor concentrations at the surfaces of the two droplets $c_S = c_{S0} e^{-(\Delta_{vap}H_m - \Delta_{vap}H_m^0)/RT}$. Under the same ambient conditions, the vapor concentration at infinity is identical for both droplets, i.e., $H_0 c_{S0} = H c_S$. Furthermore, for two droplets with the same contact angle, the dimensionless evaporation rates are equal, i.e., $\dfrac{J_{T0}}{RDc_{S0}(1-H_0)} = \dfrac{J_T}{RDc_S(1-H)}$. Combining these formulas leads to the following relationship between the total evaporation rates of the two droplets:

$$\frac{J_T}{J_{T0}} = \frac{e^{-(\Delta_{vap}H_m - \Delta_{vap}H_m^0)/RT} - H_0}{1 - H_0} \tag{7}$$

Equation (7) indicates that increasing the vaporization enthalpy or cohesive energy of a liquid reduces its evaporation rate. The calculations highlight the substantial influence of these thermodynamic parameters on evaporation dynamics, particularly in high-humidity environments (see Figure 1b). At a relative humidity of 75%, a decrease in the molar enthalpy of vaporization by as little as 0.5 kJ/mol nearly doubles the evaporation rate, while an increase in the molar enthalpy of vaporization by the same magnitude is predicted to reduce the evaporation rate by approximately 75%.

**2.5 Experimental Validation**

To validate the theoretical equations, experimental observations on the evaporation of sessile water droplets under electric fields were carried out. Deionized water droplets with a volume of 3 ± 0.3 μL were carefully deposited at the center of a horizontal electric field generated between two parallel copper sheet electrodes separated by a distance of 5 mm. The electric field strength was adjustable from 0 to 80 V/mm. During the evaporation process, side-view images of droplets were recorded by using a CMOS camera and the evaporation rates of droplets were calculated from the recorded images.

Throughout the entire pinning stage, all droplets maintained a spherical-cap shape, indicating that the applied electric field did not alter the profile of the droplets. The total evaporation rate of liquid droplets in the absence of the electric field was experimentally measured and numerically calculated. These two values were found to be almost identical, implying that the liquid evaporation is diffusion-controlled and that the presence of electrodes has little effect on the evaporation. The experimental results clearly demonstrate that the applied electric field has a pronounced effect on the water evaporation (see Figure 2b). As the electric field strength increases, a progressive reduction in the evaporation rate was observed. Under an electric field of 80V/mm, the evaporation rate is reduced by up to approximately 40% compared to the case with no electric field.

Then, molecular dynamics simulations (MDS) were performed using Materials Studio to investigate the behavior of water molecules under homogeneous electric fields ranging from 0 to 80V/mm. A cubic simulation box with side lengths of approximately 30 Å was constructed, containing a total of 700 water molecules. Water-water interaction was modeled by using the widely adopted simple point charge/extension (SPC/E) model [22]. Van der Waals interactions were described by 9-6 Lennard-Jones (LJ) potentials, while electrostatic interaction was modeled based on Coulomb's law [23,24]. Periodic boundary conditions were imposed for all three directions and a time step length of 1 fs was used. Initially, all the molecules were randomly distributed within the simulation box, and initial velocities of atoms were assigned according to the Maxwell-Boltzmann distributions at the temperature of 298 K. The system was simulated in an NVT ensemble without applying the electric field for 200,000 time steps (i.e., 200 ps) to achieve equilibrium. Following this, the electric field was applied, and the system was simulated for another 200 ps to reach equilibrium again. The average cohesive energy over the last 100 ps was taken as the equilibrium cohesive energy of water under the applied electric field (see Figure 2a).

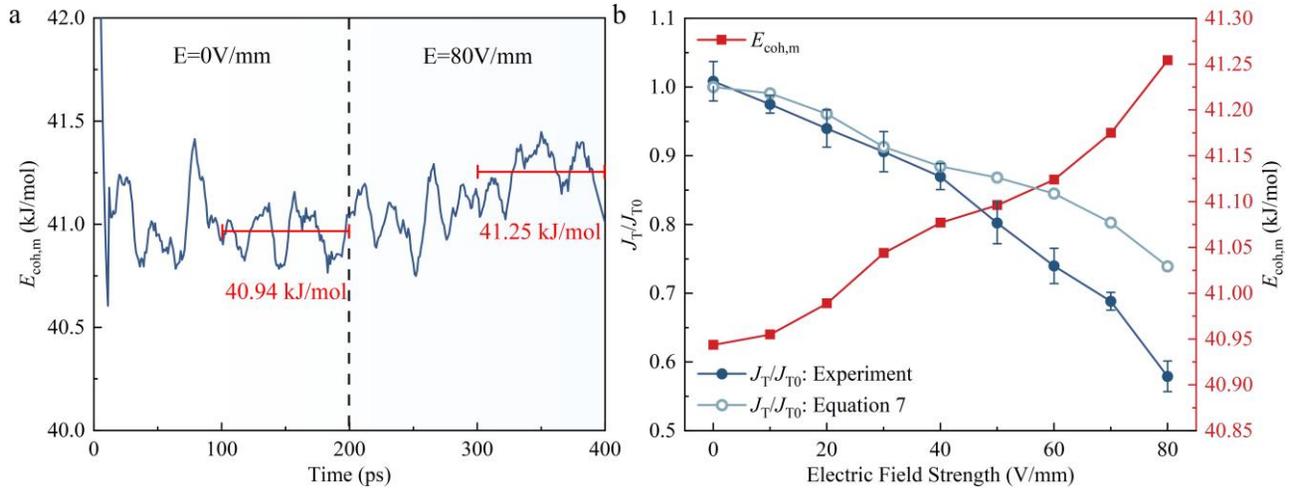

**Figure 2. (a)** Variations in the molar cohesive energy of water system with the simulation time. **(b)** Variations of the molar cohesive energy of water system and the normalized evaporation rate of water droplets as functions of electric field strength. Experimental data points show mean values from three independent measurements, with error bars indicating the standard deviation.

The simulations reveal that the cohesive energy or the vaporization enthalpy of water increases gradually with increasing electric field strength (see Figure 2b). Compared to the case without an electric field, an increase of approximately 0.3 kJ/mol in the molar vaporization enthalpy was observed under an electric field of 80 V/mm. By substituting the molar vaporization enthalpies obtained from MDS and the relative humidity measured in the ambient atmosphere into equation (7), the normalized evaporation rates $J_T/J_{T0}$, which is defined as the ratio of the total evaporation rate $J_T$ under an electric field strength of $E$ to the corresponding zero-electric field value $J_{T0}$, were calculated. The results show good agreement with the experimental measurements (see Figure 2b), thereby supporting the validity of the proposed theoretical equations. The discrepancies between the theoretical predictions and the experimental observations may result from the distortion of the electric field within the liquid droplet and the small number of atoms in the simulation system compared to that in an actual droplet.

## 3. Conclusions

In this work, the change in phase transition enthalpy of system under an electric field during an infinitesimal, isothermal, and isobaric process was first analyzed and found to be essentially equal to the non-expansion work performed on the system by the electric field. By incorporating this non-expansion work, the Clapeyron equation and the Clausius–Clapeyron equation were extended to include the variation in phase transition enthalpy. Then, by devising a reversible two-stage process, analytical expressions were derived for the variations in the saturation vapor pressure of liquids and the total evaporation rate of sessile liquid droplets with respect to changes in the molar vaporization enthalpy. The equations show that a decrease in the vaporization enthalpy of a liquid leads to an increase in its saturation vapor pressure and consequently an increase in its evaporation rate. For water, an approximately 1% change in phase transition enthalpy can cause a change of about 20% in vapor pressure and an almost twofold change in the evaporation rate.

Furthermore, experimental observations on the evaporation of sessile water droplets under electric fields were carried out. The experimental results show good agreement with the theoretical predictions, thereby supporting the validity of the proposed analytical equations. Because the energy transferred as heat to the system when applying electrical fields was neglected in the derivation presented in this paper, the proposed equations are applicable only to materials for which the electrothermal effect is insignificant. However, despite its simple origin and limitations, the results presented here may serve as an attempt to understand thoroughly how changes in phase transition enthalpy influences phase equilibrium.

## Acknowledgements

This research was supported by the National Natural Science Foundation of China (52375166).